# Predicting Different Adhesive Regimens of Circulating Particles at Blood Capillary Walls

A. Coclite[a, b], H. Mollica[a], S. Ranaldo[c, b], G. Pascazio[c, b], M. D. de Tullio[c, b], P. Decuzzi[a, *]

[a] Laboratory of Nanotechnology for Precision Medicine, nPMed, Fondazione Istituto Italiano di Tecnologia, Via Morego 30-16163, Genova, Italy.

[b] Centro di Eccellenza in Meccanica Computazionale, CEMeC, Politecnico di Bari, Via Re David, 200-70125, Bari, Italy.

[c] Dipartimento di Meccanica, Matematica e Management, DMMM, Politecnico di Bari, Via Re David, 200-70125, Bari, Italy.

[*]corresponding author: Paolo Decuzzi; paolo.decuzzi@iit.it

**ABSTRACT**

A fundamental step in the rational design of vascular targeted particles is the firm adhesion at the blood vessel walls. Here, a combined Lattice Boltzmann – Immersed Boundary model is presented for predicting the near wall dynamics of circulating particles. A moving least squares algorithm is used to reconstruct the forcing term accounting for the immersed particle, whereas ligand-receptor binding at the particle-wall interface is described via forward and reverse probability distributions. First, it is demonstrated that the model predicts with good accuracy the rolling velocity of tumor cells over an endothelial layer in a microfluidic channel. Then, particle-wall interactions are systematically analyzed in terms of particle geometries (circular, elliptical with aspect ratios 2 and 3); surface ligand densities (0.3; 0.5; 0.7 and 0.9); ligand-receptor bond strengths (1 and 2); and Reynolds numbers (Re = 0.01; 0.1 and 1.0). Depending on these conditions, four different particle-wall interaction regimens are identified, namely not adhering, rolling, sliding and firmly adhering particles. The proposed computational strategy can be efficiently used for predicting the near wall dynamics of particles with arbitrary geometries and surface properties and represents a fundamental tool in the rational design of particles for the specific delivery of therapeutic and imaging agents.

# INTRODUCTION

A plethora of nano/micro-particles have been developed over the last decade for the precise delivery of therapeutic and imaging agents in the treatment and early detection of a variety of diseases, including cancer and cardiovascular.[1, 2] Over free drug molecules and contrast agents, systemically injectable particles offer multiple advantages, such as improved organ biodistribution, enhanced accumulation at diseased sites; and protection of the therapeutic cargo from a rapid enzymatic degradation.[3-5] Top-down fabrication approaches have been proposed for precisely and independently tailoring the size, shape, surface properties and, more recently, the mechanical stiffness of particles – the so called 4S parameters in the rational design of particles.[6-9] Specifically, the particle size may vary from a few tens of nanometers to few a microns; the shape can be spherical, discoidal, cylindrical, and spheroidal; the surface can be decorated with a variety of ligand molecules for specific cell recognition; and the particle structure can be soft as cells or stiff as metals. The ability to finely tune the 4S parameters allows us, on one hand, to fabricate particles with a large variety of configurations, *de facto* enabling rational particle design, but, on the other hand, requires sophisticated computational tools for wisely selecting optimal particle configurations, depending on the biological target. Indeed, given the number of possible combinations, a rational selection solely based on experimental testing is practically unfeasible.

Moved by this need, in recent years, the authors and other scientists have started developing and employing new mathematical and computational tools for predicting the vascular and extravascular behavior of particles in terms of the 4S parameters[10]. For instance, at the macro-vascular scale, the Isogeometric Analysis (IA) was fruitfully exploited to predict the vascular deposition of micro-particles, directly infused via a catheter positioned within the left coronary artery, as a function of the endothelial receptor densities.[11, 12] Similarly, direct numerical simulations (DNS) and the immersed boundary (IB) were employed to predict the fluid–structure interaction of bodies with arbitrary shapes immersed in an incompressible fluid.[13, 14] At the microscopic scale, the immersed finite element method (IFEM) was also used to study the transport of micro and nanoparticles within whole blood and demonstrate that sub-micro and micron-sized particles would tend to be pushed laterally towards the vessel walls by the fast moving and more abundant red blood cells. [15, 16]

More recently, the Lattice Boltzmann (LB) method was also employed to solve transport problems of biological relevance. Because of its simple implementation and high parallel performance, LB is a suitable method for describing complex flow behaviors across a wide range of length and temporal scales.[17-19] This computational tool was efficiently applied to follow the dynamics of rigid particles and deformable capsules, such us red blood cells (RBCs) and leukocytes, in whole blood capillary flow. Specifically, it was applied to finely tune the geometry and viscoelastic properties of RBCs in order to accurately replicate the rheological response of whole blood as well as reduce computing burden[20-22]; predict the clustering of RBCs and microcapsules in narrow capillaries[23]; explain the role of RBCs on the vascular rolling of leukocytes[24]; determine numerically the size of the cell free layer developing next to the vessel walls[25]; and model the vascular transport of micro/nano-particles.[13, 26-28]

In this work, a LB-IB method is further developed for predicting the adhesive interaction of particles with blood vessel walls, under capillary flow. The particle surface is decorated with ligand molecules, mediating specific adhesive interaction with counter-molecules (receptors) distributed over the vessel walls. These interfacial molecular adhesive forces are computed through a probabilistic approach determining bond formation and destruction over the entire particle surface.[29] The near-wall dynamics of circular and elliptical particles, with two aspect ratios, is analyzed at three Reynolds numbers (Re=0.01, 0.1, 1.0), for three different densities of the surface ligands and two different values of the ligand-receptor chemical affinity. A direct comparison between computational predictions and experimental measurements is also presented in the case of rolling tumor cells on vascular endothelium in a microfluidic chip for assessing the model accuracy. Then, particle-wall interaction maps are derived in term of particle shape, ligand density, bond strength and flow conditions.

**COMPUTATIONAL METHOD**

The mathematical method used to model the fluid evolution and the fluid-structure interaction, proposed and validated by Coclite and colleagues [13], is briefly described in the following.

***The combined lattice Boltzmann immersed boundary (LB-IB) method.*** The evolution of the fluid is defined in terms of a set of N discrete distribution functions $\{f_i\}$(i=0,..., N−1) which obey the dimensionless Boltzmann equation,

$$f_i(\boldsymbol{x} + \boldsymbol{e}_i \Delta t, t + \Delta t) - f_i(\boldsymbol{x}, t) = -\frac{\Delta t}{\tau}[f_i(\boldsymbol{x},t) - f_i^{eq}(\boldsymbol{x},t)], \qquad (1)$$

in which **x** and t are the spatial and time coordinates, respectively; [$e_i$](i=0,...,N−1) is the set of discrete velocities; $\Delta t$ is the time step; and $\tau$ is the relaxation time given by the unique non-null eigenvalue of the collision term in the BGK-approximation [30]. The kinematic viscosity of the flow is related to the single relaxation time $\tau$ as $v = c_s^2 \left(\tau - \frac{1}{2}\right) \Delta t$ being $c_s = \frac{1}{\sqrt{3}} \frac{\Delta x}{\Delta t}$ the reticular speed of sound. The moments of the distribution functions define the fluid density $\rho = \sum_i f_i$, velocity $\boldsymbol{u} = \sum_i f_i \boldsymbol{e}_i / \rho$, and the pressure $p = c_s^2 \rho = c_s^2 \sum_i f_i$. The local equilibrium density functions [$f_i^{eq}$] (i=0,...,N−1) are expressed by the Maxwell-Boltzmann (MB) distribution,

$$f_i^{eq}(\boldsymbol{x},t) = \omega_i \rho \left[1 + \frac{1}{c_s^2}(\boldsymbol{e}_i \cdot \boldsymbol{u}) + \frac{1}{2c_s^4}(\boldsymbol{e}_i \cdot \boldsymbol{u})^2 - \frac{1}{2c_s^2}\boldsymbol{u}^2\right]. \qquad (2)$$

On the two-dimensional square lattice with N = 9 speeds (D2Q9) [31], the set of discrete velocities is given by:

$$\boldsymbol{e}_i = \begin{cases} (0,0), & if\ i = 0 \\ \left(\cos\left(\frac{(i-1)\pi}{2}\right), \sin\left(\frac{(i-1)\pi}{2}\right)\right), & if\ i = 1-4 \\ \sqrt{2}\left(\cos\left(\frac{(2i-9)\pi}{4}\right), \sin\left(\frac{(2i-9)\pi}{4}\right)\right), & if\ i = 5-8 \end{cases}, \qquad (3)$$

with the weight, $\omega_i = 1/9$ for i = 1−4, $\omega_i = 1/36$ for i = 5−8, and $\omega_0 = 4/9$. Here, we adopt a discretization in the velocity space of the MB distribution based on the Hermite polynomial expansion of this distribution [32].

An effective forcing term accounting for the boundary presence, $\mathcal{F}_i$, can be included as an additional factor on the right-hand side of eq.(1).

Following the argument from Guo et al.[33], also developed in [34-37], $\mathcal{F}_i$ is given by:

$$\mathcal{F}_i = \left(1 - \frac{1}{2\tau}\right) \omega_i \left[\frac{e_i - u}{c_s^2} + \frac{e_i \cdot u}{c_s^4} e_i\right] \cdot f_{ib}, \quad (4)$$

where $f_{ib}$ is the body force term evaluated through the formulation by Favier et al. [38], combined with the moving least squares reconstruction [39] in the immersed boundary technique by Coclite et al. [13]. Due to the presence of the forcing term $\mathcal{F}_i$, the macroscopic quantities, given by the moments of the distribution functions, are obtained as:

$$\rho = \sum_i f_i, \quad (5)$$

$$\rho u = \sum_i f_i e_i + \frac{\Delta t}{2} \mathcal{F}_i, \quad (6)$$

It is proved that in such a framework one can recover the forced Navier–Stokes equations with second order accuracy [33, 38]. In the present model the forcing term accounts for the presence of an arbitrary shaped body into the flow-field, whereas the external boundaries of the computational domain are treated with the known-velocity bounce back conditions by Zou and He [40].

***Pressure and viscous stresses.*** Let $nl$ be the number of linear elements composing the surface of the immersed body being $l$ the element index, the pressure and viscous stresses exerted by the immersed body are:

$$F_p(t) = \sum_{l=1}^{nl}(-p_l\ \boldsymbol{n}_l) S_l, \qquad (7)$$

$$F_p(t) = \sum_{l=1}^{nl}(\bar{\tau}_l \cdot \boldsymbol{n}_l) S_l, \qquad (8)$$

where $\bar{\tau}_l$ and $p_l$ are the viscous stress tensor and the pressure evaluated in the centroid of the $l$-th element, respectively; $\boldsymbol{n}_l$ is the outward normal unit vector while $S_l$ is the length of the $l$-th element. The pressure and velocity derivatives in eq.s (7) and (8) are evaluated considering a probe in the normal positive direction of each element, the probe length being $1.2\Delta x$, and using the moving least squares formulation cited [13]. In this framework, the velocity derivatives evaluated at the probe are considered equal to the ones on the linear element centroid as previously done by the authors. [13, 14]

***Wall-particle interaction.*** The adhesion model used in the present work is based on the works by Sun et al.[24, 29]. Ligand and receptor molecules are distributed over the particle and vessel wall surfaces, respectively. Ligand molecules are modeled as linear springs which, by interacting with wall receptor molecules, tend to establish bonds (ligand-receptors bonds) and support a mechanical force $f_{l,b}$ given as:

$$\boldsymbol{f}_{l,b} = \sigma(y_l - y_{cr,eq})\boldsymbol{n}_l, \qquad (9)$$

with $y_l$ the bond length, $y_{cr,eq}$ the equilibrium bond length and $\sigma$ the spring constant. The receptor density is assumed uniform; the solid wall is supposed completely covered by receptive molecules. In the present model, all springs have the same spring constant, $\sigma$. The total adhesive force, $\boldsymbol{F}_b$, is obtained by integrating $\boldsymbol{f}_{l,b}$ over the particle perimeter. Bonds can be only generated if the minimum separation distance between the particle boundary and the wall is smaller than a critical value, $y_{cr} = 6.8 \times 10^{-3} H$. The equilibrium bond length, resulting in a null force, is chosen as $y_{cr,eq} = 0.5\ y_{cr}$. All lengths are normalized by the channel height, $H$. The linear spring constant is computed in *lattice units* and non-dimensionalized through the term, $\frac{\varrho_{ref} v^2}{H}$, where $\rho_{ref}$, $H$, and $v$ are reference density, length and kinematic viscosity, respectively.

The bond formation is regulated by a forward probability function,

$$P_f = 1 - \exp(-k_f Nl \Delta t), \qquad (10)$$

with $k_f$ forward bond rate and $Nl$ the number of ligand actually probing the surface (number of *active* elements) over the total number of linear elements. At each time step, a pre-existing bond can be destroyed according to the reverse probability function,

$$P_r = 1 - \exp\left(-k_{r0} exp\left(\frac{(\sigma-\sigma^*)(y_l-y_{cr,eq})^2}{2k_B T}\right)\Delta t\right). \qquad (11)$$

Here, $k_{r0}$ is the reverse bond rate, $\sigma^*$ is the equilibrium spring constant (taken as 0.5 $\sigma$), and $k_B T$ is the thermal potential. [24, 29] The equilibrium spring constant $\sigma^*$ enables to model two different classes of ligand-receptor bonds: 'slip' bonds for $\sigma > \sigma^*$, where forces exerted on the bond facilitate disentanglement; 'catch' bonds for $\sigma < \sigma^*$, where forces exerted on the bond facilitate entanglement. Here, by fixing $\sigma^* = 0.5\sigma$, slip bonds are considered which are far more common in the case of leukocyte and cancer cell rolling/adhesion. [41]

A Van der Waals like potential is implemented to model the particle-wall interaction. The force **F**w is so applied along the solid walls positive normal directions into the particle centroid[24],

$$F_w = \frac{H_k}{8\sqrt{2}}\sqrt{\frac{r}{\epsilon^5}}\,n, \qquad (12)$$

being $n$ the solid wall normal direction unit vector, $H_k$ the Hamacker constant, $r$ the particle radius and $\epsilon$ the separation distance between the particle and the wall. The Hamacker constant is non-dimensionalized through the term, $\rho_{ref}Hv^2$.

*Fluid-structure interaction strategy.* The total force *F(t)* and total moment *M(t)* acting on the immersed body are evaluated in time and the translation and rotation of the particle are updated at each Newtonian dynamics time step by an explicit second order scheme. Therefore, the linear and angular accelerations are obtained directly as:

$$\dot{u}(t) = F(t)/m, \qquad (13)$$

$$\dot{\omega}(t) = M(t)/I, \qquad (14)$$

being *m* and *I* the particle mass and inertia moment of the two-dimensional particle about its centroid, respectively. The linear and angular velocities are computed as:

$$u(t) = \frac{2}{3}\left(2u(t-\Delta t) - \frac{1}{2}u(t-2\Delta t) + \dot{u}(t)\Delta t\right), \qquad (15)$$

$$\omega(t) = \frac{2}{3}\left(2\omega(t-\Delta t) - \frac{1}{2}\omega(t-2\Delta t) + \dot{\omega}(t)\Delta t\right), \qquad (16)$$

with $\Delta x = \Delta t = 1$. Here, a weak coupling approach between the fluid and the particle is implemented. Note that this approach is unconditionally stable for small velocity variations [42], which is indeed the case of the present work.

## RESULTS

**Cell rolling in a capillary flow.** To reproduce typical capillary flow conditions, a single channel microfluidic chip is realized in PDMS, following standard fabrication procedures (**Figure.1a**).[43] First, a negative mold of the channel is generated, upon UV light cross-linking, baking and development of a SU-8 film. Then, a PDMS replica of the mold is realized and peeled off after curing. Two circular holes of ~ 1 mm are punched into the PDMS layer, constituting the inlet and outlet of the microfluidic chip. Finally, following an oxygen plasma treatment, the PDMS layer is bonded to a glass slide. The resulting microfluidic chip has a channel with width W = 210 μm, height H = 42 μm and length L = 2.70 cm. Top and side views as well as an optical microscopy image of the chip are presented in **Figure.1a**.

To establish a confluent cell layer resembling the microvascular endothelium, human umbilical vein endothelial cells (HUVECs) are cultured within the channel. Specifically, after autoclaving the chip in DI water for two hours at 120°C, the channel is first filled with fibronectin (20 μg/mL) and then with HUVECs ($2\times10^6$ cells/mL). Then, the chip is kept in an incubator for about 2 days, until a confluent monolayer of HUVECs is established. At this point, $10^6$ colon rectal tumor cells (HCT-15) are injected in the endothelialized channel via a syringe pump at different flow rates, namely Q = 50, 100, 150 and 200 nL/min. The vascular transport of HCT-15 cells is monitored for about 15 minutes using a fluorescent microscope and the rolling velocity is estimated within the region of interest. Note that the considered flow rates Q correspond to physiological and tumor characteristic wall shear rates S, namely S = 13.5, 27, 40.5 and 54 s$^{-1}$ ($S =$ $6Q/(WH^2)$). Also, the corresponding mean flow velocities U are 94.48, 188.9, 283.4 and 377.9 µm/s ($U = Q/(WH)$), respectively.

Still images of cells rolling over the endothelial monolayer within the chamber are presented in **Figure.1b**, at different time points. The rolling velocity $u_r$ is defined as the ratio between the distance traveled by the cell and the period of observation and is derived by post processing the fluorescent microscopy images. As shown in **Figure.1c**, the rolling velocity (red dots) increases linearly as the flow rate Q grows ($R^2 = 0.966$), ranging from 113.9 ± 4.13 for 50 nL/min to 322.2 ± 22.8 for 200 nL/min. This was expected and confirmed by theoretical and numerical predictions.

The theoretical rolling velocity $u_{th}$ is derived assuming the cell as a rigid sphere of diameter d rolling in a rectangular channel pushed by a flow with rate Q, as derived for a channel of rectangular cross section [44], so that

$$u_{th} = \frac{3}{2}\frac{Q}{WH}\left[1 - \left(1 - \frac{d}{H}\right)^2\right]. \qquad (17)$$

Considering that the diameter of HCT-15 cells ranges between 14 and 20 µm (d = 15 ± 3 µm), it results $u_{th}$ = 83.15 ± 19.7, 166.1 ± 39.4, 249.0 ± 59.1 and 333.0 ± 78.7 µm/s, for each of the four considered flow rates. This rolling velocity is also estimated using the present LB – IB model assuming the cell as a circular particle, settled at a distance $3\times10^{-3}$ H from the wall and with a ligand density 0.3. The assumed ligand density value of 0.3 returns a good agreement between the experimental and numerical predictions for the cell rolling velocity over four different flow rates. The ligand density represents the ratio between the number of ligand molecules on the circulating cell and the number of receptor molecules distributed over the endothelial cells lining the blood vessel walls. As expected, both the numerical and theoretical rolling velocities vary linearly with Q and are in very good agreement with each other ($R^2$ = 0.994). The experimental, theoretical and numerical rolling velocities are all plotted in **Figure.1c**, for the four flow rates Q. Given the variation in cell diameter, shadowed areas are used to present the rolling velocities, whose upper and lower limits are associated with the bigger and smaller cell diameters, respectively. The present LB – IB model accurately predicts the rolling velocity of tumor cells over a wide range of flow rates.

**Modeling the adhesion dynamics of near-wall circulating particles.** In this section, the adhesive dynamics of particles circulating in close proximity of the blood vessel walls is predicted employing the present LB – IB computational approach. Vascular adhesion is assessed in terms of physiological parameters, such as the local hemodynamic conditions – the Reynolds number ($Re = \frac{H \, u_{max}}{v_{ref}}$), based on the upper wall velocity, $u_{max}$, the capillary height, $H$, and the reference kinematic viscosity, $v_{ref}$; and particle parameters, such as the particle shape – circular and elliptical; and density of ligand molecules ($\rho_l$) decorating the particle perimeter.

The computational domain resembles the near-wall region in a capillary flow and is limited at the bottom (y = 0) by a fixed wall (the vessel wall in a blood capillary) and at the top (y = H) by a moving wall (interface between the cell-free layer and the core of the blood capillary) (**Figure.2a**). Within this domain, a linear shear rate is imposed and the upper wall has a velocity $u_{max}$. The height H of the computational domain coincides with the height of the microfluidic chip used before for the experimental validation. The computational domain is confined within the area $[0, 10H] \times [0, H]$, where H is discretized with 200 points. Periodic boundary conditions are imposed on the two sides (x = 0 and x = 10H); zero slip velocities are imposed at the bottom (u = 0) and top (u = $u_{max}$) walls so that the linear flow field follows the relationship $u_x(0, y) = u_{max} y/H = (Re \cdot v_{ref}/H) y/H$.

Particles, initially at rest, are placed in the computational domain at a separation distance from the bottom wall equal to $y_0 = 3 \times 10^{-3}$ H. Three different geometries are considered for the particles, namely circular, elliptical with an aspect ratio 2, and elliptical with an aspect ratio 3 (**Figure.3a**). The characteristic size is chosen as to keep constant the total area enclosed by the particle, namely A = 0.025 $H^2$. Thus, the circular particle has a diameter of 0.18 H, and the elliptical particles have axial lengths equal to 0.25H × 0.125H and 0.31H × 0.103H, respectively. On the particle perimeter, ligand molecules are uniformly distributed with a density $\rho_l$ of 0.3, 0.5, 0.7, and 0.9 (**Figure.2b-c**). On the vessel walls, a receptor density equal to 1 is imposed. The ligand-receptor bonds are characterized by an adhesive bond strength σ and a biochemical affinity $k_f/k_{r,0} = 8.5 \times 10^3$. All parameters used in the model are listed in **Table 1**, with their dimensional and non-dimensional values, and the schematic representations of the computational domain and particles are given in **Figure.2**.

**Vascular adhesion dynamics for circular particles.** Data on the adhesive dynamics of circular particles are shown in **Figure.3** and listed in **Table.2**, for Re = 0.01, 0.1 and 1.0. A circular particle, initially settled in close proximity of the vessel wall ($y_0 = 3.0 \times 10^{-3}$ H) (**Figure.3a**), rapidly forms ligand-receptor bonds initiating the adhesion process. Note that the initial separation distance $y_0$ is smaller than the critical distance for bond formation ($y_{cr} = 6.8 \times 10^{-3}$ H; dashed lines in **Figure.3**).

The equilibrium position of the particle with respect to the wall is given by $y_{min}$, the minimum separation distance, which is plotted in **Figure.3b** and **3f**, respectively for Re = 0.1 and 1.0. The particle with a ligand density $\rho_l = 0.3$ moves away from the wall returning an equilibrium separation distance $y_{min} = 7.6 \times 10^{-3}$ H, which is larger than the critical distance for bond formation. Thus, for $\rho_l = 0.3$ and smaller, the ligand density is insufficient to induce the formation of any stable bonds and the particle moves away from the wall – *not adhering particle*. For larger ligand densities, the separation distance at equilibrium reduces returning values of $6.67 \times 10^{-3}$ H, $6.29 \times 10^{-3}$ H and $6.3 \times 10^{-3}$ H, for $\rho_l = 0.5$, 0.7, and 0.9 respectively. These are all cases where the equilibrium position is smaller than the critical bond distance $y_{cr}$ and stable ligand-receptor bonds are formed. Indeed, as $\rho_l$ increases, the hydrodynamic forces exerted over the particle are redistributed over a larger number of ligands thus diminishing the deformation of each ligand-receptor bonds and moving the particle closer to the wall (**Figure.3b** and **3f**).

As documented in **Figure.3c** and **3g**, the percentage of active ligands increases with $\rho_l$, in other words the number of closed ligand-receptor bonds grows with the number of ligands decorating the particle perimeter. Note that, since ligand-receptor binding is defined in a statistical manner, the number of bonds oscillates over time around an average value. For $\rho_l = 0.3$, the number of active bonds is zero (*not adhering particle*), whereas it grows to 0.020, 0.036 and 0.048, for $\rho_l = 0.5$, 0.7 and 0.9, respectively. Oscillations in the number of active ligands appears as bands in **Figure.3c** and **3g**. It is here important to highlight that, due to the small region of contact between a circular particle and the wall, only a small number of ligand-receptor bonds are formed even in the case of high ligand densities. As from **Figure.3c** and **3g**, the percentage of active ligands is equal to 2% for $\rho_l = 0.5$ and grows only up to ~ 5% for $\rho_l = 0.9$.

The kinematic parameters, namely the angular rotation $\theta$ and longitudinal velocity $u_x$ of the particle, are presented in the remaining insets of **Figure.3**. The variation of $\theta$ over time is given in **Figure.3d** and **3h**. It shows a steady and linear increase of $\theta$, thus implying a constant angular velocity $\Omega$ of the particle over the wall – *rotating, not adhering particle*. The rotational velocity reduces as the number of ligand-receptor bonds increases and is equal to $\Omega H/u_{max}$ = 0.29, 0.25, 0.125, and 0.125, respectively for $\rho_l$ = 0.3, 0.5, 0.7 and 0.9, at Re = 0.1. For larger Reynolds numbers (Re=1.0), the angular velocity $\Omega$ exhibits a negligible variation with $\rho_l$, possibly because of the larger hydrodynamic dislodging forces. In this condition, $\Omega H/u_{max}$ is equal to 0.29, 0.284, 0.280, and 0.280.

Finally, the normalized longitudinal velocity $u_x/u_{max}$ of the particle is plotted versus time in **Figure.3e** and **3i**. Even for this physical quantity, oscillations appear around an average value, following what has been already reported for the number of active ligands. Oscillations are larger for the smaller Reynolds numbers. Indeed, for Re =0.1, the average longitudinal velocity is nearly zero in the case of $\rho_l = 0.7$ and 0.9, and grows up to $4.08\times10^{-2}$ for $\rho_l = 0.5$. For $\rho_l = 0.3$, the normalized longitudinal velocity is $4.80\times10^{-2}$ with no oscillation in that the particle is not adhering to the wall and travels as a rigid body passively transported by the blood flow. For Re = 1.0, oscillations are smaller and the longitudinal velocity higher due to the larger hydrodynamic dislodging forces. Numerical values for all displacement and kinematic parameters are listed in **Table.2** for ease of comparison. **Table.2** shows also that for Re = 0.01, all particles exhibiting a ligand density $\rho_l \geq 0.5$ have zero rotational and longitudinal velocity, implying that these particles can form stable bonds with the wall – *firmly adhering particles*. Differently, particles with $\rho_l < 0.5$ roll without adhering to the wall – *rolling, not adhering particles*.

**Vascular adhesion dynamics for elliptical particles.** Data on the adhesive dynamics of elliptical particles are shown in **Figure.4** and listed in **Table.3**, for Re = 0.01, 0.1 and 1.0 and for aspect ratios equal to 2 and 3. The elliptical particle, initially settled in close proximity of the vessel wall ($y_0 = 3.0\times10^{-3}$ H) and with its major axes pointing orthogonally to the wall, rapidly forms ligand-receptor bonds initiating the adhesion process (**Figure.4a**). Note that the separation distance $y_0$ is smaller than the critical distance for bond formation ($y_{cr} = 6.8\times10^{-3}$ H).

At low Reynolds number, the adhesion dynamics of elliptical particles is qualitatively similar to that of circular particles. As shown in **Figure.4b**, the equilibrium position $y_{min}$ is rapidly reached and preserved for the whole simulation period. At first, an abrupt variation in $y_{min}$ is observed, which is related to the initial orientation of the particle with respect to the flow field and its sudden rotation. Also, as compared with circular particles, the equilibrium position $y_{min}$ is slightly higher for a given ligand density $\rho_l$. This could possibly be ascribed to higher hydrodynamic forces exerted over elliptical particles. Very differently, at Re = 1.0 and for sufficiently low ligand densities ($\rho_l = 0.3$), firm deposition of elliptical particles on the wall is impaired and the dislodging forces are strong enough to induce a periodic particle rotation over the wall – *rolling, not adhering particle*. This is shown in **Figure.4f** where $y_{min}/H$ oscillates and stays constant (transient adhesion), only for a small portion of the observation time.

Furthermore, the number of closed ligand-receptor bonds is larger for elliptical particles at all given $\rho_l$, but for $\rho_l=0.3$ (**Figure.4c** and **4g**). Indeed, elliptical particles expose a larger portion of their perimeter to the wall allowing for a larger percentage of ligands to be engaged with their counter-molecules (receptors) on the wall (> 2-fold). Also note that, for $\rho_l = 0.3$, the number of closed ligand-receptor bonds is equal to zero for both circular and elliptical particles.

The angular rotation $\theta$ is plotted in **Figure.4d** and **4h**. For Re = 0.1 (**Figure.4d**), particles move from the original vertical position ($\theta = 0$) and progressively deposit on the wall tending to the more stable configuration $\theta = \pi/2$. This rotation occurs quite abruptly for $\rho_l$ larger than 0.3. Differently, for Re = 1.0 (**Figure.4h**), the rolling and not adhering particle ($\rho_l=0.3$ and Re = 1.0) shows a continuously growing $\theta$ with spikes in angular velocities Ω (local derivative of θ with respect to time) corresponding to a quasi-vertical position of the particle. For larger ligand densities, θ reaches the steady state value of θ = π/2 implying that the particle does not rotate anymore after laying down on the wall.

Finally, the normalized longitudinal velocity $u_x/u_{max}$ is plotted in **Figure.4e** and **4i**. For all considered cases, the velocity is not zero but constant for the whole observation period beside for the rolling and not adhering particle (**Figure.4i**). The not zero velocity implies that the not rotating elliptical particles, once deposited horizontally over the wall, tend to slide longitudinally breaking old bonds at the trailing edge, forming new bonds at the leading edge and along the particle body – *sliding, not adhering particles*. Indeed, the larger is the number of active ligands and the lower is the sliding velocity of the particle.

Numerical values for all displacement and kinematic parameters are listed in **Table.3,** for an aspect ratio 2, and **Table.4**, for an aspect ratio 3**,** for ease of comparison. **Table.3** and **Table.4** show also that for Re = 0.01, all particles exhibiting a ligand density $\rho_l \geq$ 0.5 have zero rotational and longitudinal velocity, implying that these particles can form stable bonds with the wall – *firmly adhering particles*. Differently, particles with $\rho_l <$ 0.5 roll without adhering to the wall – *rolling, not adhering particles*.

**Particle-wall interaction regimens.** As described in the previous paragraphs, depending on the flow and particle properties, different regimens of particle-wall interaction can be documented: firmly adhering, rolling, sliding and not adhering particles. This is summarized in **Figure.5** and **Figure.6**, where the rolling velocity $u_{roll}/u_{max}$ and probability of adhesion $P_a$ are presented as a function of the considered three different shapes – circular, elliptical with aspect ratio 2 and 3; ligand densities $\rho_l$ – ranging from 0.3 to 0.9; Reynolds numbers – varying from 0.01 to 1.0; and bond strength σ – equal to 1 (soft bond) and 2 (rigid bond). **Figure.5** presents a contour plot for the normalized rolling velocity $u_{roll}/u_{max}$, whereas **Figure.6** gives a contour plot for the probability of adhesion, $P_a$. This quantity is defined as the ratio between the number of active bonds and the maximum number of bonds that can be closed at any given time during the adhesion process and represents the likelihood of forming stable bonds at the particle-wall interface. The maximum number of bonds is readily calculated as a function of the particles geometry and orientation with respect to the wall. Both physical quantities ($u_r$ and $P_a$) are affected in a similar fashion by the flow and particle properties. Specifically, low Reynolds numbers and high ligand densities (upper-left area) are associated with zero rolling velocities and *firmly adhering particles*. Indeed, under these conditions, the hydrodynamic dislodging forces are moderately low (low Re) and are readily balanced by the high adhesive interactions (high $\rho_l$). At the other extreme, high Reynolds numbers and low ligand densities (lower-right area) are associated with *not adhering particles*. Under these conditions, the hydrodynamic dislodging forces (high Re) cannot be balanced by the adhesive interactions (low $\rho_l$). In between these two limiting conditions, particles are observed to move relatively to the substrate. With circular particles and elliptical particles at moderate $\rho_l$, continuously rolling over the wall is documented. On the other hand, with elliptical particles at high $\rho_l$, longitudinal sliding over the wall is observed. Note that rolling of elliptical particles is limited by their larger rotational inertia. However, longer bonds may facilitate rolling and slender particles as depicted in **Figure.7**. Finally, adhesion is favored by stronger bonds in that, for fixed dislodging forces, higher $\sigma$ are associated with lower ligand-receptor bond energies ($\propto F_b^2/\sigma$).

## CONCLUSIONS AND FUTURE PERSPECTIVES

A combined Lattice Boltzmann-Immersed Boundary (LB-IB) model was developed for predicting the adhesive interactions of circulating particles with walls lining a blood vessel. Particles were decorated with ligand molecules forming molecular bonds with counter molecules (receptors) uniformly distributed over the wall. Three different particle shapes were considered (circular and elliptical with aspect ratio 2 and 3) and transported in a linear laminar flow, characterized by physiologically relevant Reynolds numbers (from 0.01 to 1.0).

First, the computational model was validated by estimating the velocities of quasi-circular cells rolling over a continuous endothelial layer in a microfluidic chip. For different values of the Reynolds number, predictions from the LB – IB model were in good agreement with experimental data, thus confirming the accuracy of the proposed approach. Then, the interaction of circular and elliptical particles with the wall was studied varying systematically the particle shape, ligand density, ligand-receptor bond strength and flow conditions. As a function of the above independent parameters, particle-wall interaction maps were derived documenting four possible regimens: firmly adhering, sliding, rolling, and not adhering particles.

The proposed LB-IB model can be accurately employed to predict the vascular dynamics and adhesion interactions of systemically injected particles. Relevant biophysical parameters can be efficiently modulated allowing for systematic analyses and supporting the rational design of particles for drug delivery and imaging.

**Contributors.** The study was designed by AC and PD. AC developed and integrated the LB code with the IB code. HM performed all microfluidic experiments. SR developed the MPI parallel scheme. GP and MdT developed the IB model. PD coordinated the project. Results were analyzed and discussed by all authors.

**Funding**. This project was partially supported by the European Research Council, under the European Union's Seventh Framework Programme (FP7/2007-2013)/ERC grant agreement no. 616695; AIRC (Italian Association for Cancer Research) under the individual investigator grant no. 17664; the Italian Institute of Technology.

# REFERENCES

1. Peer, D., et al., *Nanocarriers as an emerging platform for cancer therapy.* Nat Nanotechnol, 2007. **2**(12): p. 751-60.

2. Mulder, W.J., et al., *Imaging and nanomedicine in inflammatory atherosclerosis.* Sci Transl Med, 2014. **6**(239): p. 239sr1.

3. Bao, G., S. Mitragotri, and S. Tong, *Multifunctional nanoparticles for drug delivery and molecular imaging.* Annu Rev Biomed Eng, 2013. **15**: p. 253-82.

4. Min, Y., et al., *Clinical Translation of Nanomedicine.* Chem Rev, 2015. **115**(19): p. 11147-90.

5. Muthu, M.S., et al., *Nanotheranostics - application and further development of nanomedicine strategies for advanced theranostics.* Theranostics, 2014. **4**(6): p. 660-77.

6. Key, J., et al., *Soft Discoidal Polymeric Nanoconstructs Resist Macrophage Uptake and Enhance Vascular Targeting in Tumors.* ACS Nano, 2015. **9**(12): p. 11628-41.

7. Key, J., et al., *Engineering discoidal polymeric nanoconstructs with enhanced magneto-optical properties for tumor imaging.* Biomaterials, 2013. **34**(21): p. 5402-10.

8. Euliss, L.E., et al., *Imparting size, shape, and composition control of materials for nanomedicine.* Chem Soc Rev, 2006. **35**(11): p. 1095-104.

9. Anselmo, A.C. and S. Mitragotri, *Impact of particle elasticity on particle-based drug delivery systems.* Advanced Drug Delivery Reviews, 2017. **108**(Supplement C): p. 51-67.

10. Decuzzi, P., *Facilitating the Clinical Integration of Nanomedicines: The Roles of Theoretical and Computational Scientists.* ACS Nano, 2016. **10**(9): p. 8133-8.

11. Hossain, S.S., et al., *In silico vascular modeling for personalized nanoparticle delivery.* Nanomedicine (Lond), 2013. **8**(3): p. 343-57.

12. Hossain, S.S., T.J. Hughes, and P. Decuzzi, *Vascular deposition patterns for nanoparticles in an inflamed patient-specific arterial tree.* Biomech Model Mechanobiol, 2014. **13**(3): p. 585-97.

13. Coclite, A., et al., *A combined Lattice Boltzmann and Immersed boundary approach for predicting the vascular transport of differently shaped particles.* Computers & Fluids, 2016. **136**: p. 260-271.

14. de Tullio, M. and G. Pascazio, *A moving-least-squares immersed boundary method for simulating the fluid–structure interaction of elastic bodies with arbitrary thickness.* Journal of Computational Physics, 2016. **325**: p. 201-225.

15. Lee, T.-R., et al., *On the near-wall accumulation of injectable particles in the microcirculation: smaller is not better.* Scientific reports, 2013. **3**.

16. Lee, T.R., et al., *Quantifying uncertainties in the microvascular transport of nanoparticles.* Biomech Model Mechanobiol, 2014. **13**(3): p. 515-26.

17. Succi, S., *Lattice Boltzmann across scales: from turbulence to DNA translocation.* European Physical Journal B, 2008. **64**(3-4): p. 471-479.

18. Succi, S., *The lattice Boltzmann equation: for fluid dynamics and beyond*. 2001: Oxford university press.

19. Aidun, C.K. and J.R. Clausen, *Lattice-Boltzmann Method for Complex Flows.* Annual

Review of Fluid Mechanics, 2010. **42**: p. 439-472.

20. Fedosov, D.A., B. Caswell, and G.E. Karniadakis, *A Multiscale Red Blood Cell Model with Accurate Mechanics, Rheology, and Dynamics.* Biophysical Journal, 2010. **98**(10): p. 2215-2225.

21. Sun, C.H. and L.L. Munn, *Particulate nature of blood determines macroscopic rheology: A 2-D lattice Boltzmann analysis.* Biophysical Journal, 2005. **88**(3): p. 1635-1645.

22. Kruger, T., F. Varnik, and D. Raabe, *Efficient and accurate simulations of deformable particles immersed in a fluid using a combined immersed boundary lattice Boltzmann finite element method.* Computers & Mathematics with Applications, 2011. **61**(12): p. 3485-3505.

23. McWhirter, J.L., H. Noguchi, and G. Gompper, *Flow-induced clustering and alignment of vesicles and red blood cells in microcapillaries.* Proceedings of the National Academy of Sciences of the United States of America, 2009. **106**(15): p. 6039-6043.

24. Sun, C., C. Migliorini, and L.L. Munn, *Red blood cells initiate leukocyte rolling in postcapillary expansions: a lattice Boltzmann analysis.* Biophys J, 2003. **85**(1): p. 208-22.

25. Fedosov, D.A., et al., *Blood flow and cell-free layer in microvessels.* Microcirculation, 2010. **17**(8): p. 615-28.

26. Gekle, S., *Strongly Accelerated Margination of Active Particles in Blood Flow.* Biophys J, 2016. **110**(2): p. 514-20.

27. Tan, J.F., et al., *Characterization of Nanoparticle Dispersion in Red Blood Cell*

*Suspension by the Lattice Boltzmann-Immersed Boundary Method.* Nanomaterials, 2016. **6**(2).

28. Basagaoglu, H., et al., *Two- and three-dimensional lattice Boltzmann simulations of particle migration in microchannels.* Microfluidics and Nanofluidics, 2013. **15**(6): p. 785-796.

29. Sun, C. and L.L. Munn, *Lattice Boltzmann simulation of blood flow in digitized vessel networks.* Comput Math Appl, 2008. **55**(7): p. 1594-1600.

30. Bhatnagar, P.L., E.P. Gross, and M. Krook, *A Model for Collision Processes in Gases. I. Small Amplitude Processes in Charged and Neutral One-Component Systems.* Phys. Rev., 1954. **94**(3): p. 511-525.

31. Qian, Y.H., D. Dhumieres, and P. Lallemand, *Lattice Bgk Models for Navier-Stokes Equation.* Europhysics Letters, 1992. **17**(6bis): p. 479-484.

32. Shan, X.W., X.F. Yuan, and H.D. Chen, *Kinetic theory representation of hydrodynamics: a way beyond the Navier-Stokes equation.* Journal of Fluid Mechanics, 2006. **550**: p. 413-441.

33. Guo, Z., C. Zheng, and B. Shi, *Discrete lattice effects on the forcing term in the lattice Boltzmann method.* Phys Rev E Stat Nonlin Soft Matter Phys, 2002. **65**(4 Pt 2B): p. 046308.

34. De Rosis, A., S. Ubertini, and F. Ubertini, *A partitioned approach for two-dimensional fluid-structure interaction problems by a coupled lattice Boltzmann-finite element method with immersed boundary.* Journal of Fluids and Structures, 2014. **45**: p. 202-215.

35. De Rosis, A., S. Ubertini, and F. Ubertini, *A Comparison Between the Interpolated Bounce-Back Scheme and the Immersed Boundary Method to Treat Solid Boundary Conditions for Laminar Flows in the Lattice Boltzmann Framework.* Journal of Scientific Computing, 2014. **61**(3): p. 477-489.

36. Suzuki, K., K. Minami, and T. Inamuro, *Lift and thrust generation by a butterfly-like flapping wing-body model: immersed boundary-lattice Boltzmann simulations.* Journal of Fluid Mechanics, 2015. **767**: p. 659-695.

37. Wang, Y., et al., *An immersed boundary-lattice Boltzmann flux solver and its applications to fluid-structure interaction problems.* Journal of Fluids and Structures, 2015. **54**: p. 440-465.

38. Favier, J., A. Revell, and A. Pinelli, *A Lattice Boltzmann-Immersed Boundary method to simulate the fluid interaction with moving and slender flexible objects.* Journal of Computational Physics, 2014. **261**: p. 145-161.

39. Vanella, M. and E. Balaras, *A moving-least-squares reconstruction for embedded-boundary formulations.* Journal of Computational Physics, 2009. **228**(18): p. 6617-6628.

40. Zou, Q.S. and X.Y. He, *On pressure and velocity boundary conditions for the lattice Boltzmann BGK model.* Physics of Fluids, 1997. **9**(6): p. 1591-1598.

41. Marshall, B.T., et al., *Direct observation of catch bonds involving cell-adhesion molecules.* Nature, 2003. **423**(6936): p. 190.

42. Zhang, Q. and T. Hisada, *Studies of the strong coupling and weak coupling methods in FSI analysis.* International Journal for Numerical Methods in Engineering, 2004.

**60**(12): p. 2013-2029.

43. Manneschi, C., et al., *A microfluidic platform with permeable walls for the analysis of vascular and extravascular mass transport.* Microfluidics and Nanofluidics, 2016. **20**(8): p. 113.

44. Bird, R.B., *Transport phenomena.* Applied Mechanics Reviews, 2002. **55**(1): p. R1-R4.

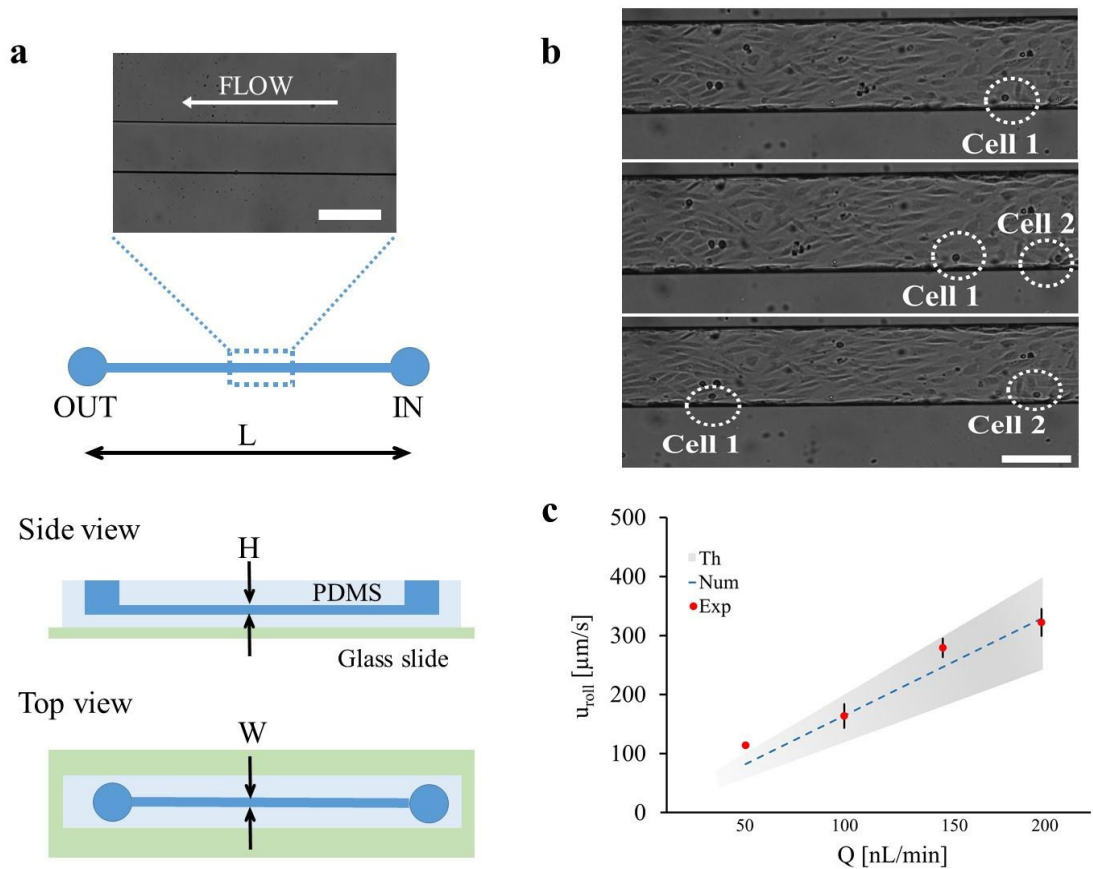

**Figure 1: HCT-15 cells rolling on an HUVEC monolayer into a single channel microfluidic chip. a.** Schematic representation of the single channel microfluidic chip with definition of the main geometric quantities. From top to bottom: brightfield epi-fluorescent microscope image of the region of interests (scale bar 250 μm); side and top views of the chip (L=2.7 cm, H=42 μm, W=210 μm). **b.** Representative images of HCT-15 cells rolling over a confluent monolayer of HUVECs ($10x$ magnification, scale bar

250 μm). **c.** Rolling velocity of HCT-15 under four different flow conditions (50, 100, 150, and 200 nL/min) estimated via numerical and theoretical analyses.

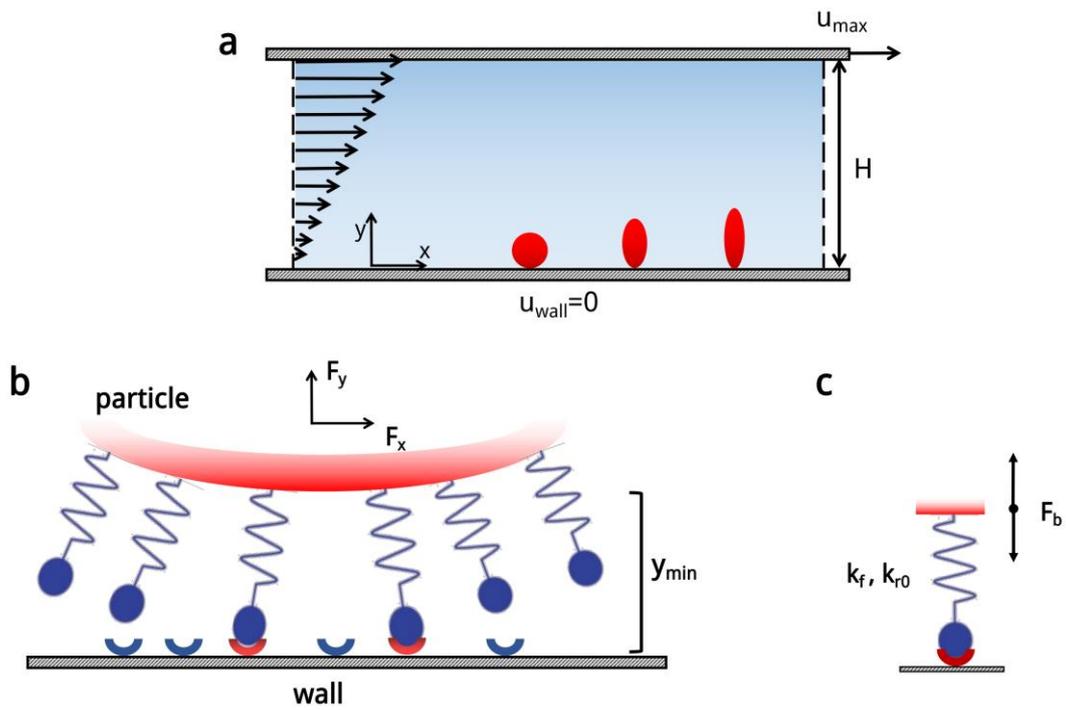

**Figure 2: Particle transport in a linear laminar flow. a.** Schematic representation of the computational domain. **b.** Ligand distributed over the particle perimeter interacting with receptors distributed over the vessel wall. **c.** Ligand-receptor bond modeled as a spring with characteristic forward $k_f$ and reverse $k_{r0}$ strengths.

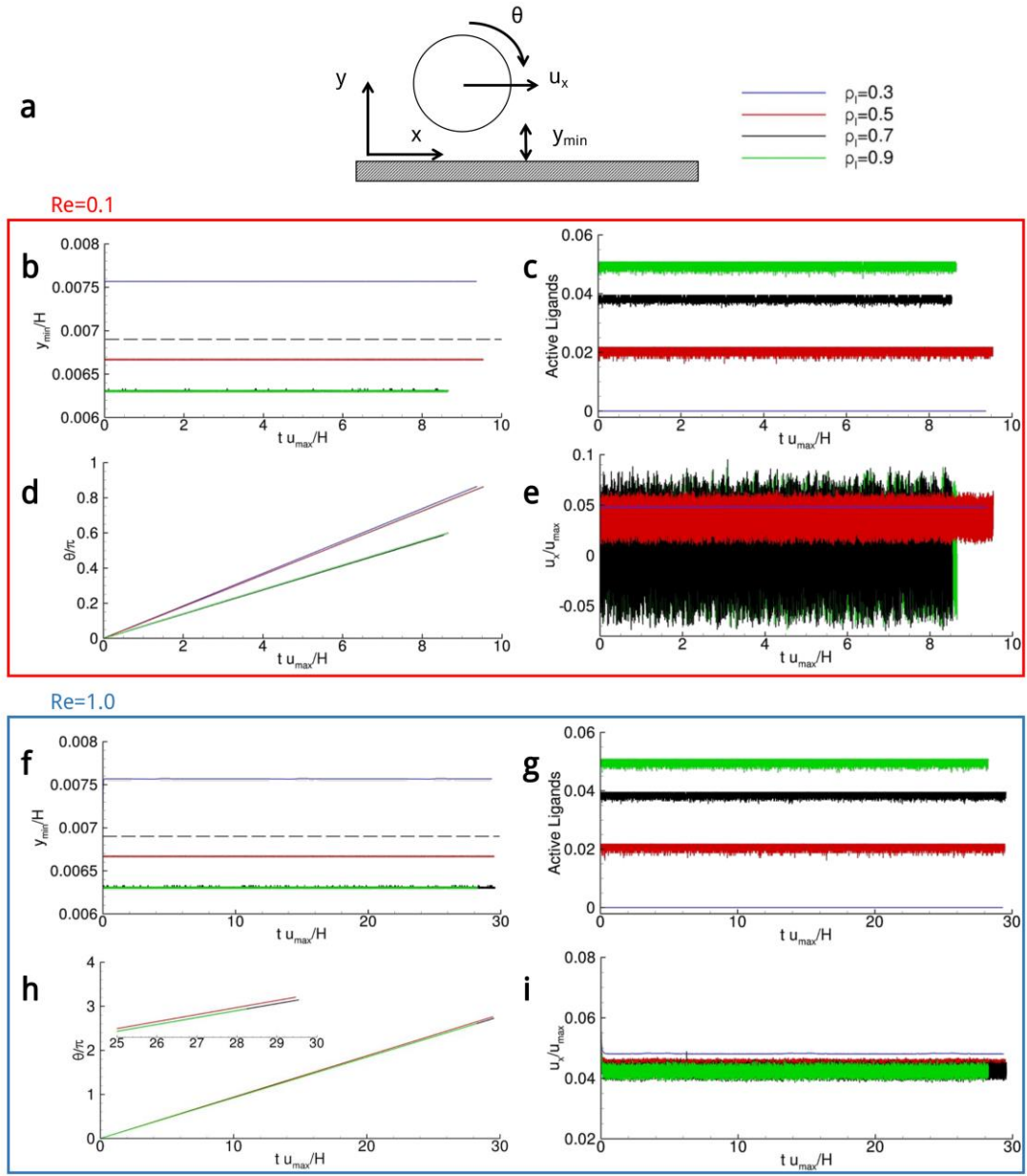

**Figure 3**: **Vascular adhesion of circular particles ( $\sigma = 2$ ). a.** Schematic representation of the problem. **(b, f)** Particle separation distance from the wall versus time. The dashed line corresponds to $y_{cr}$. **(c, g)** Active over total number of ligands versus time. **(d, h)** Angular rotation, $\theta$, versus time where the inset presents a magnified view within the interval $25 \leq tu_{max} \leq 30$. **(e, i)** Normalized rolling velocity versus time.

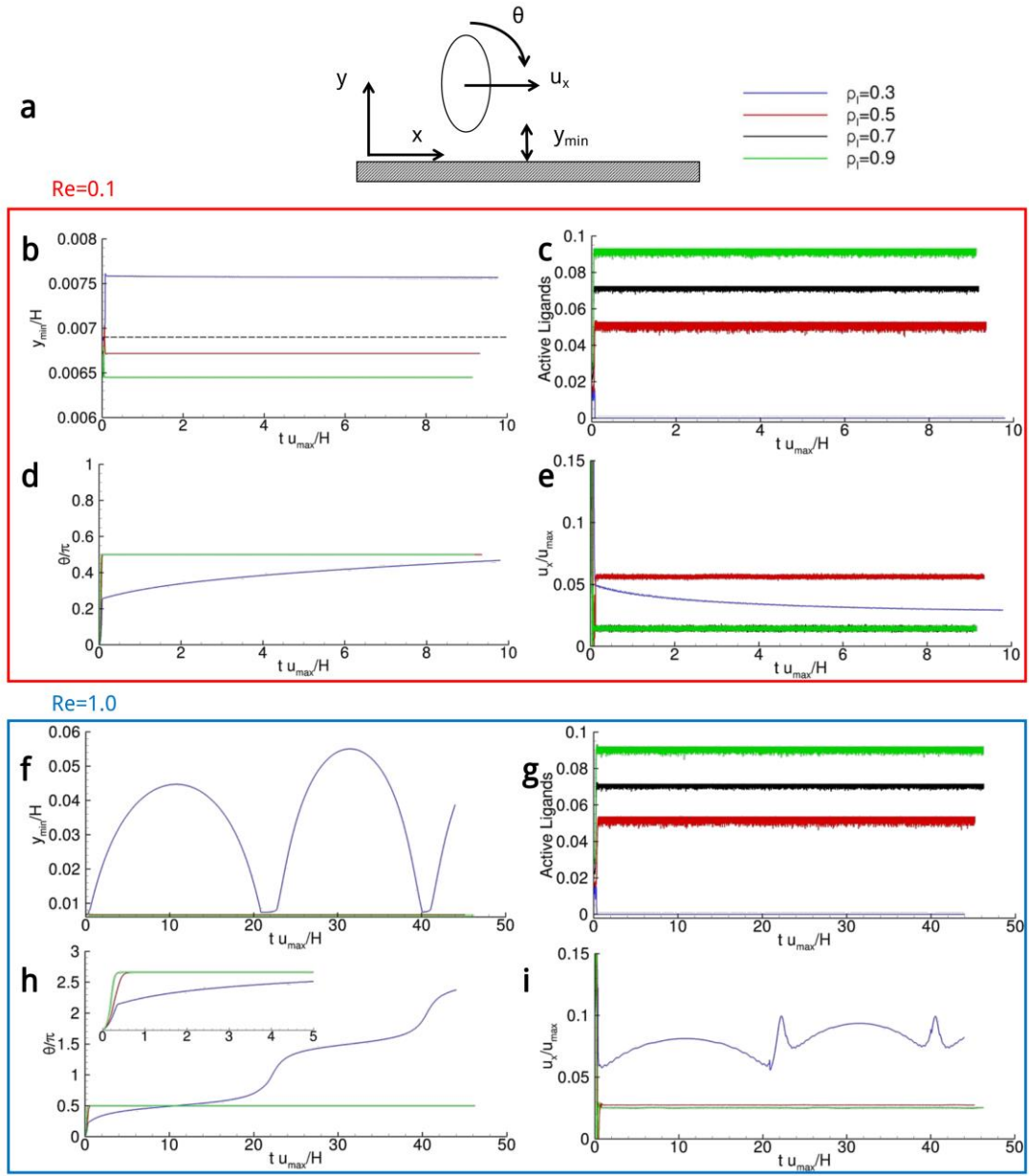

**Figure 4**: **Vascular adhesion of elliptical particles ($\sigma = 2$). a.** Schematic representation of the problem. **(b, f)** Particle separation distance from the wall versus time. The dashed line corresponds to $y_{cr}$. **(c, g)** Active over total number of ligands versus time. **(d, h)** Angular rotation, $\theta$, versus time where the inset presents a magnified view within the interval $0 \leq tu_{max} \leq 5$. **(e, i)** Normalized rolling velocity versus time.

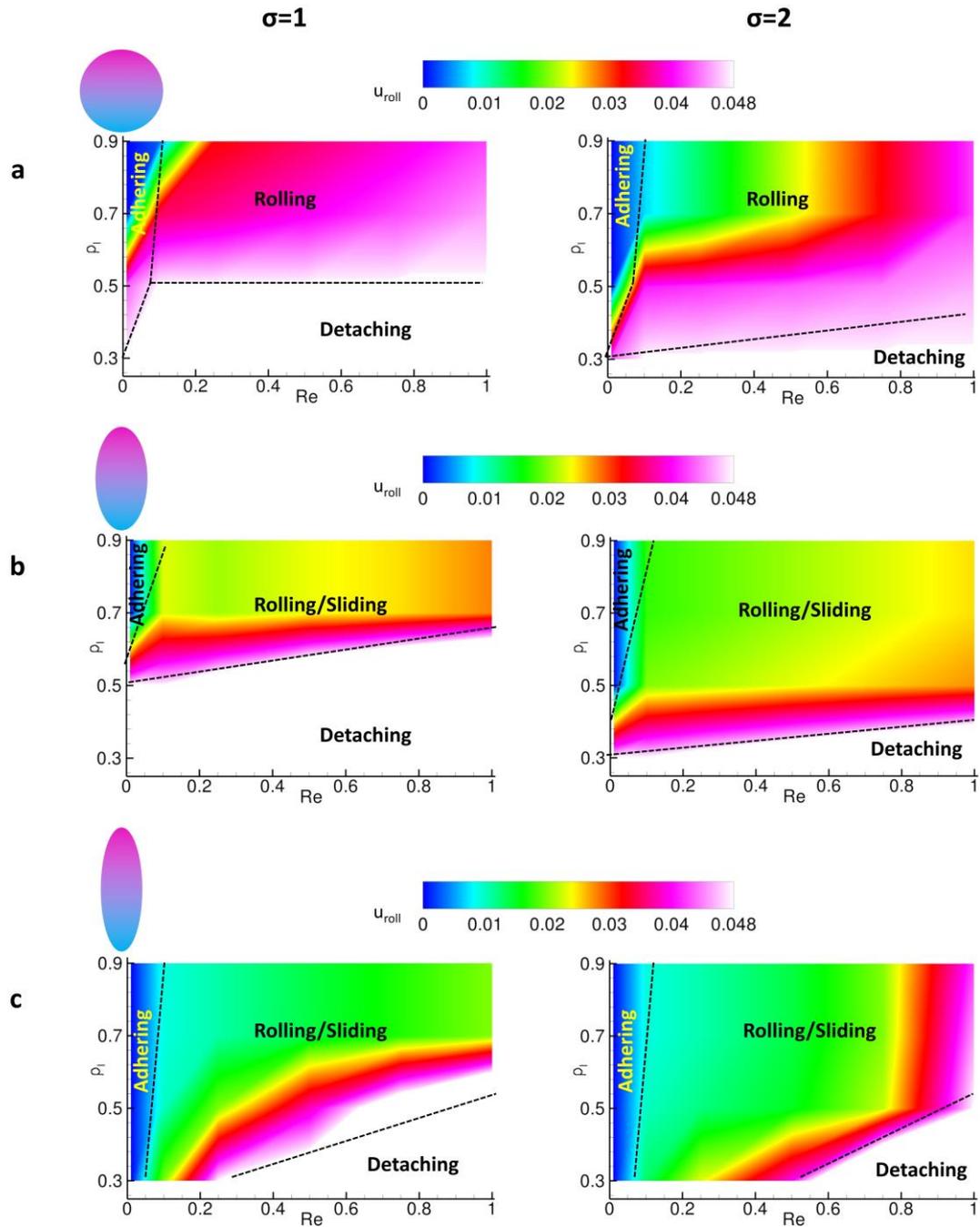

**Figure 5: Contour plots for the rolling velocity. a.** Circular particle transport with soft ($\sigma = 1$) and rigid ($\sigma = 2$) ligand-receptor bonds. **b.** Elliptical particle, with aspect ratio 2, transport with soft ($\sigma = 1$) and rigid ($\sigma = 2$) ligand-receptor bonds. **c.** Elliptical particle, with aspect ratio 3, transport with soft ($\sigma = 1$) and rigid ($\sigma = 2$) ligand-receptor bonds.

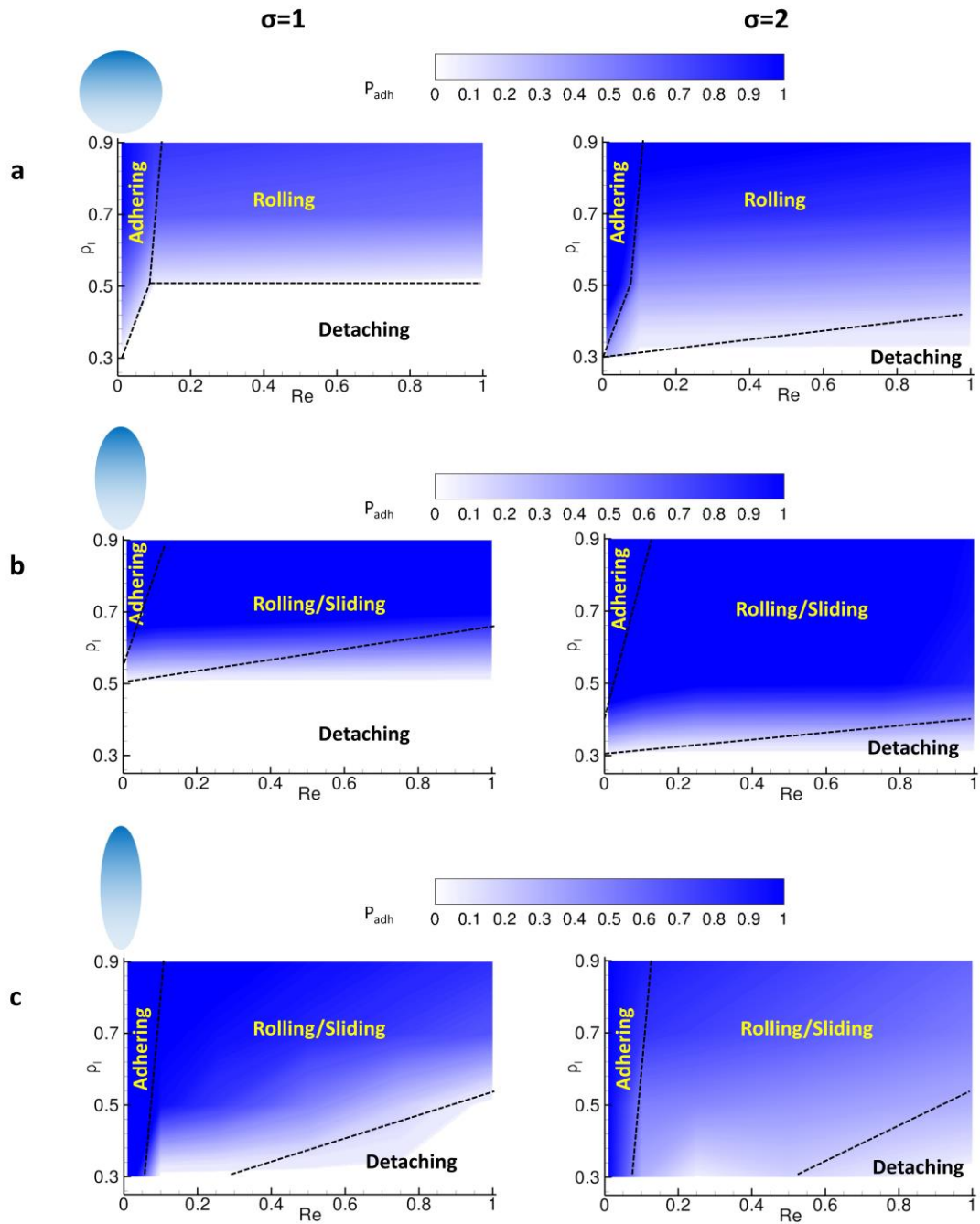

**Figure 6: Contour plots for the probability of adhesion. a.** Circular particle transport with soft ($\sigma = 1$) and rigid ($\sigma = 2$) ligand-receptor bonds. **b.** Elliptical particle, with aspect ratio 2, transport with soft ($\sigma = 1$) and rigid ($\sigma = 2$) ligand-receptor bonds. **c.** Elliptical particle, with aspect ratio 3, transport with soft ($\sigma = 1$) and rigid ($\sigma = 2$) ligand-receptor bonds.

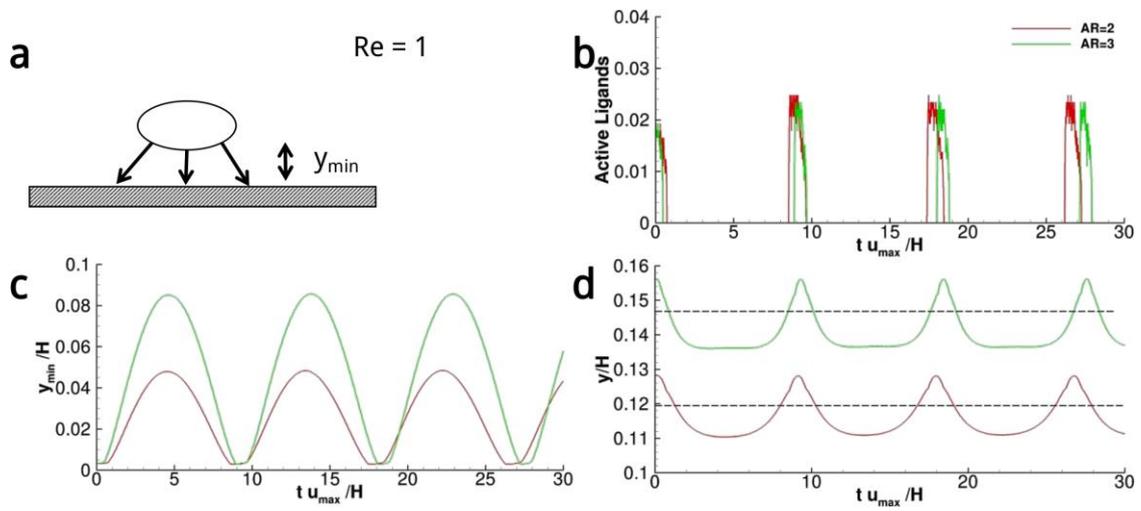

**Figure 7**: **Vascular transport of elliptical particles with different critical bond length. a.** Schematic representation of the problem. **b.** Active over total number of ligands versus time. **c.** Particle separation distance from the wall versus time. **d.** Centroid lateral position versus time.

| | Parameters | SI-Unit Value | Lattice-Unit Value | Dimensionless Value | Dimensionless Group |
|---|---|---|---|---|---|
| **Reference Variables** | | | | | |
| $H$ | Channel thickness | 40 μm | 200 | 1 | - |
| $\rho_{ref}$ | Density of water | $10^3$ kg/m$^3$ | 1 | 1 | - |
| $v_{ref}$ | Kinematic viscosity of water | $1.2 \times 10^{-6}$ m$^2$/s | 1/6 | 1 | - |
| **Dimensionless quantities** | | | | | |
| $d$ | Particles diameter | 7.20 μm  10.0 μm  12.4 μm | 36 (AR=1)  50 (AR=2)  62 (AR=3) | 0.18  0.25  0.31 | $\dfrac{d}{H}$ |
| $u_{max}$ | Top-wall velocity | $3\times10^{-4}$ m/s  $3\times10^{-3}$ m/s  $3\times10^{-2}$ m/s | $0.8\times10^{-5}$  $0.8\times10^{-4}$  $0.8\times10^{-3}$ | 0.01  0.1  1 | $Re = \dfrac{u_{max}H}{v_{ref}}$ |
| $y_{cr}$ | Critical bond length | 0.050 μm | 1.30 | $6.8\times10^{-3}$ | $\dfrac{y_{cr}}{H}$ |
| $y_{cr,eq}$ | Equilibrium bond length | 0.025 μm | 0.65 | $3.4\times10^{-3}$ | $\dfrac{y_{cr}}{H}$ |
| $H_k$ | Hamacker constant | $1.0\times10^{-21}$ J | $0.1\times10^{-8}$ | $1.74\times10^{-8}$ | $\dfrac{H_k}{\rho_{ref}v_{ref}^2 H}$ |
| $\sigma$ | Spring constant | $1.0\times10^{-3}$ N/m  $2.0\times10^{-3}$ N/m | 1.0  2.0 | 27.8  55.5 | $\dfrac{\sigma}{\rho_{ref}v_{ref}^2/H}$ |
| $\rho_l$ | Ligand density | - | 0.3  0.5  0.7  0.9 | | - |

**Table 1:** Parameters used in the computational experiments expressed in the SI-unit system and in lattice-unit system along with their dimensionless groups. Note that, the channel thickness, $H$; the water density, $\rho_{ref}$; the water kinematic viscosity, $v_{ref}$ are used throughout the formulation to present in dimensionless form all other dependent physical quantities while all quantities for the description of the physical problem are shaded in green.

| 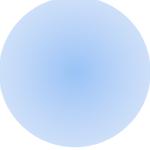 | | Ligand density | | | |
|---|---|---|---|---|---|
| | | 0.3 | 0.5 | 0.7 | 0.9 |
| Reynolds number | 0.01 $y_{min}/H$ | $7.57 \times 10^{-3}$ | $6.0 \times 10^{-3}$ | $6.0 \times 10^{-3}$ | $6.0 \times 10^{-3}$ |
| | 0.01 $u_{roll}/u_{max}$ | $4.43 \times 10^{-2}$ | 0 | 0 | 0 |
| | 0.01 $\Omega H/u_{max}$ | 0.280 | 0 | 0 | 0 |
| | 0.1 $y_{min}/H$ | $7.57 \times 10^{-3}$ | $6.67 \times 10^{-3}$ | $6.29 \times 10^{-3}$ | $6.29 \times 10^{-3}$ |
| | 0.1 $u_{roll}/u_{max}$ | $4.80 \times 10^{-2}$ | $4.08 \times 10^{-2}$ | $0.69 \times 10^{-2}$ | $0.68 \times 10^{-2}$ |
| | 0.1 $\Omega H/u_{max}$ | 0.29 | 0.25 | 0.125 | 0.125 |
| | 1 $y_{min}/H$ | $7.57 \times 10^{-3}$ | $6.67 \times 10^{-3}$ | $6.29 \times 10^{-3}$ | $6.29 \times 10^{-3}$ |
| | 1 $u_{roll}/u_{max}$ | $4.80 \times 10^{-2}$ | $4.55 \times 10^{-2}$ | $4.20 \times 10^{-2}$ | $4.20 \times 10^{-2}$ |
| | 1 $\Omega H/u_{max}$ | 0.29 | 0.284 | 0.280 | 0.280 |

**Table 2:** Circular particle kinematics and dynamics quantities obtained for σ=2. Separation distance, $y_{min}/H$, rolling velocity, $u_{roll}/u_{max}$, and rotational velocity, $\Omega H/u_{max}$, are tabulated as function of the Reynolds number and the density of ligands ($\rho_l$).

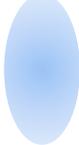

|  |  | Ligand density | | | |
|---|---|---|---|---|---|
|  |  | 0.3 | 0.5 | 0.7 | 0.9 |
| Reynolds number / 0.01 | $y_{min}/H$ | $8.2 \times 10^{-3}$ | $6.1 \times 10^{-3}$ | $6.1 \times 10^{-3}$ | $6.1 \times 10^{-3}$ |
|  | $u_{roll}/u_{max}$ | $5.00 \times 10^{-2}$ | 0 | 0 | 0 |
|  | $\Omega H/u_{max}$ | 0.1 | 0 | 0 | 0 |
| 0.1 | $y_{min}/H$ | $7.76 \times 10^{-3}$ | $6.71 \times 10^{-3}$ | $6.47 \times 10^{-3}$ | $6.47 \times 10^{-3}$ |
|  | $u_{roll}/u_{max}$ | $2.88 \times 10^{-2}$ | $5.58 \times 10^{-2}$ | $1.51 \times 10^{-2}$ | $1.40 \times 10^{-2}$ |
|  | $\Omega H/u_{max}$ | 0.05 | 0 | 0 | 0 |
| 1 | $y_{min}/H$ | $27,3 \times 10^{-3}$ | $6.71 \times 10^{-3}$ | $6.47 \times 10^{-3}$ | $6.47 \times 10^{-3}$ |
|  | $u_{roll}/u_{max}$ | $7.81 \times 10^{-2}$ | $2.75 \times 10^{-2}$ | $2.52 \times 10^{-2}$ | $2.52 \times 10^{-2}$ |
|  | $\Omega H/u_{max}$ | 0.40 | 0 | 0 | 0 |

**Table 3:** Elliptical particle, with aspect ratio 2, kinematics and dynamics quantities obtained for σ=2. Separation distance, $y_{min}/H$, rolling velocity, $u_{roll}/u_{max}$, and rotational velocity, $\Omega H/u_{max}$, are tabulated as function of the Reynolds number and the density of ligands ($\rho_l$).

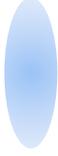

|  |  | Ligand density | | | |
|---|---|---|---|---|---|
|  |  | 0.3 | 0.5 | 0.7 | 0.9 |
| Reynolds number | 0.01 | | | | |
| | $y_{min}/H$ | $6.1 \times 10^{-3}$ | $6.1 \times 10^{-3}$ | $6.1 \times 10^{-3}$ | $6.1 \times 10^{-3}$ |
| | $u_{roll}/u_{max}$ | 0 | 0 | 0 | 0 |
| | $\Omega H/u_{max}$ | 0 | 0 | 0 | 0 |
| | 0.1 | | | | |
| | $y_{min}/H$ | $7.59 \times 10^{-3}$ | $6.71 \times 10^{-3}$ | $6.47 \times 10^{-3}$ | $6.47 \times 10^{-3}$ |
| | $u_{roll}/u_{max}$ | $1.12 \times 10^{-2}$ | $2.82 \times 10^{-2}$ | $2.10 \times 10^{-2}$ | $2.10 \times 10^{-2}$ |
| | $\Omega H/u_{max}$ | 0.04 | 0 | 0 | 0 |
| | 1 | | | | |
| | $y_{min}/H$ | $27{,}3 \times 10^{-3}$ | $6.71 \times 10^{-3}$ | $6.47 \times 10^{-3}$ | $6.47 \times 10^{-3}$ |
| | $u_{roll}/u_{max}$ | $6.00 \times 10^{-2}$ | $2.70 \times 10^{-2}$ | $2.17 \times 10^{-2}$ | $2.17 \times 10^{-2}$ |
| | $\Omega H/u_{max}$ | 0.42 | 0 | 0 | 0 |

**Table 4:** Elliptical particle, with aspect ratio 3, kinematics and dynamics quantities obtained for σ=2. Separation distance, $y_{min}/H$, rolling velocity, $u_{roll}/u_{max}$, and rotational velocity, $\Omega H/u_{max}$, are tabulated as function of the Reynolds number and the density of ligands ($\rho_l$).